\documentclass[a4paper,usenatbib]{mnras}

\usepackage{newtxtext,newtxmath}

\usepackage[T1]{fontenc}
\usepackage{ae,aecompl}

\usepackage{graphicx}
\usepackage{amsmath}	
\usepackage{amssymb}

\AtBeginShipout{%
  \ifnum\value{page}>1 %
    \typeout{* Additional boxing of page `\thepage'}%
    \setbox\AtBeginShipoutBox=\hbox{\copy\AtBeginShipoutBox}%
  \fi
}

\title[Quantifying the Smoothness of the Stellar Halo]{Quantifying the Smoothness of the Stellar Halo:\\ A Link to Accretion History}

\author[L. Lancaster et al.]{
Lachlan Lancaster,$^{1,2}$
\thanks{E-mail: lachlanl@princeton.edu}
Vasily Belokurov,$^{2,3}$ N. Wyn Evans$^{2}$
\\
$^{1}$Department of Astrophysical Sciences, Princeton University, 4 Ivy Lane, 08544, Princeton, NJ, USA\\
$^{2}$Institute of Astronomy, University of Cambridge, Madingley Road, Cambridge, CB3 0HA, UK\\
$^{3}$Center for Computational Astrophysics, Flatiron Institute, 162 5th Avenue, 10010, New York, NY, USA
}

\date{Accepted XXX. Received YYY; in original form ZZZ}

\pubyear{2017}

\begin{document}
\label{firstpage}
\pagerange{\pageref{firstpage}--\pageref{lastpage}}
\maketitle

\begin{abstract}
We investigate the utility of the 3-dimensional Two-Point Correlation
Function (3D 2PCF) in quantifying substructure in the stellar halo of
the Milky Way, particularly as a means of constraining the accretion
history of our Galaxy. We use RR Lyrae variable stars from two
different surveys as tracers of the structure in the Galactic stellar
halo. We compare our measurements of the 3D 2PCF in these datasets to
a suite of simulations of the formation of the stellar halo from
Bullock and Johnston (2005).  While there is some room for
interpretation, we find that the amounts of structure to be broadly
consistent with the simulations, while appearing smoother than average
within the inner halo and at small scales. This suggests a preferred
accretion history scenario in which the Milky Way's stellar halo
acquired most of its mass about $\sim$8 Gigayears ago, and has been
largely quiescent since. Finally, we discuss the prospects of
statistical tools such as the 2PCF in the \textit{Gaia} era of
galactic archaeology.
\end{abstract}




\section{Introduction}
\label{sec:intro}

The stellar halo of the Milky Way represents an interesting testbed for theories of galaxy formation. Its old stellar population with low metallicity and $\alpha$-element enhancement developed very early in the history of the Galaxy's formation \citep{Eggen62}. Yet opinions originally differed on whether this component was a remnant of an early stage of `galactic collapse' \citep{Eggen62} or whether it consisted of tidally stripped remnants of the hierarchical structure formation of the galaxy \citep{SearleZinn78}, as predicted in the Cold Dark Matter (CDM) cosmological paradigm of structure formation \citep{Majewski96,BandJ05,Abadi06, Cooper2010}. After this debate arose, it was clearly pointed out that determination of the structure and substructure of the stellar halo would constrain its history \citep{Majewski1993}, mostly because a hierarchical structure formation process should leave behind many remnants~\citep{Johnston98,WhiteSpringel2000,Bullock2001}.

Thanks to many large surveys over the past 20 years, such as the Sloan Digital Sky Survey (SDSS) \citep{SDSS2000}, the Two Micron All Sky Survey (2MASS) \citep{2MASS2006}, the VST ATLAS \citep{Shanks2015} and the Dark Energy Survey \citep[DES][]{DES2016}, we have developed an excellent picture of the size and shape of the stellar halo as an oblate spheroid, following a stepped density profile with radius~\citep{Yanny2000,Ivezic2000,LarsenHumphreys2003,Juric08, Deason11}. This revolution in understanding the structure of the stellar halo has been accompanied by an outpouring of discoveries of its substructure, be it the tidal streams of dwarf galaxies~\citep{Ibata95, Ivezic2000, MD2001, Majewski2003, Duffau2006, FoS2006, Newberg2007,Shipp2018}, globular cluster streams~\citep{Odenkirchen2003, Belokurov2006, GD1, ATLASstream, Shipp2018}, or low galactic latitude ring-like structures
\citep{Ibata2003,Yanny2003}. While the discovery of such abundant high latitude substructure has widely confirmed the hierarchical structure formation scenario over the `collapse' of \cite{Eggen62} (whose results were also shown to be largely due to selection bias by \cite{ChibaBeers2000}), there is still some debate as to whether some fraction of the stellar halo is composed of a disrupted component of the galactic disk, whose stars formed \textit{in situ}~\citep{Newberg2002,Momany2006,Janesh2016}.

For this reason, several authors endeavored to create quantitative measures of the substructure in the stellar halo of the Milky Way~\citep{Lemon2004, NewbergYanny05, Bell2008, Starkenburg2009, Helmi2011, Deason11}. These studies have been used to place limits on the fraction of stars in the halo that are formed \textit{in situ} versus accreted through hierarchical structure formation. In parallel with these developments (and often at the same time), many have accepted the likely dominance of accreted structure in the formation of the stellar halo, and used quantitative measures to try to characterize this substructure. Starting with \cite{Lemon2004}, statistics such as count-in-cell, angular correlation functions, and the Lee2D statistic were used on stars from the Millennium Galaxy Catalog \citep{MGC03} to study substructure in the inner stellar halo ($r \lesssim 10$ kpc) and provided little evidence for significant structure beyond the Sagittarius Stream. \cite{Bell2008} extended the use of a count-in-cell like statistic to a significantly larger sample of $\sim 4\times 10^6$ main sequence turnoff (MSTO) stars from Sloan Digital Sky Surevey (SDSS) data release 5, comparing this measure of structure with the simulations (based purely on a hierarchical structure formation scenario) of \cite{BandJ05} and found broad consistency, pointing towards a substantial fraction of the stellar halo being formed through accretion. The work of \cite{Bell2008} can be compared to the analysis of \cite{Deason11}, who instead used a smaller sample of photometrically identified A-type stars (Blue Horizontal Branch or BHB, and Blue Stragglers or BS), taking advantage of the substantially reduced systematic errors in determination of distances to these tracers. According to \citet{Deason11}, the stellar halo, probed out to $r\sim 40$ kpc by the BHBs and BSs appears to be rather smooth, in tension with the earlier measurements of \cite{Bell2008}. Note that some of the discrepancy could perhaps be attributed to the differences in the choice of the stellar tracers. 

As the spectroscopic surveys started to catch up with the all-sky imaging campaigns, the analysis of substructure was expanded by including velocity information. The inclusion of this information was largely motivated by theoretical work which suggested that substructure in phase space should be much more long-lived than in configuration space~\citep[e.g.][]{Johnston96,HelmiWhite99,ReFiorentin2005}.
This velocity space analysis began with \cite{Starkenburg2009}, who used a sample of 101 K giants from the Spaghetti Survey \citep{Morrison2000} and \cite{Xue2009,Xue11}, who used BHB stars from SDSS along with a \lq 4-distance' metric on position space along with line-of-sight velocity. They again found broad consistency with a hierarchical structure formation model when compared to the \cite{BandJ05} simulations.  This analysis was then expanded upon for BHBs in SDSS using a \lq 4-distance' statistic by \cite{Cooper2011} and for MSTO stars in SDSS using a count-in-cell statistic by \cite{Helmi2011} and compared against a separate set of cosmological simulations from the Aquarius suite \citep{Cooper2010}. Again, there was broad agreement with the hierarchical structure formation paradigm, except for a decrement in structure within the inner halo.  Most recently, \cite{Janesh2016} essentially updated the work of \cite{Starkenburg2009} using a sample of 4,568 K giants, stretching to $r\sim 125$ kpc from the SDSS Sloan Extension for Galactic Understanding and Exploration (SEGUE) project. They use the \lq 4-distance' statistic on this extended set of data to argue that, while there exists evidence of substructure, most of it is likely associated with the Sagittarius (Sgr) stream and that 50\% of their sample is not associated with any substructure at all.

While these authors have quantified the amount of substructure, their studies often focused on the questions of the formation mechanism of different parts of the halo (\textit{in situ} versus accreted) and/or the statistical discovery of late-time accreted structures which are still relatively bound. Here, instead, we choose to use similar statistical techniques to measure the faint signatures of the early-time accretion history of the Milky Way's stellar halo. To this end, we assume that all of the Galactic halo's RR Lyrae have been contributed via accretion. We base this assumption on the fact that in-situ formation models require a substantial net rotation in the stellar halo \citep[see e.g.][]{Mccarthy2012, Tissera2018}, while none has been observed in the RR Lyrae sample presently~\citep{Deason17}. We appeal to this property to use the RR Lyrae variables as tracers of the \textit{accreted substructure} of the stellar halo in order to constrain this accretion history, via comparison to numerical simulations of this accretion process. It may seem that the choice of the RR Lyrae as a halo tracer could inflict biases in our study. We note that, while the fraction of RR Lyrae does indeed vary substantially in the Milky Way satellites, all of the Galactic dwarfs without exception contain a share of these horizontal branch pulsators~\citep[see][for details]{Sesar2014,Fiorentino2015}.

The structure of this paper is as follows.  In Section \ref{sec:methods}, we review the basic properties of the statistical tool that we use in this work: the 3D spatial Two-Point Correlation Function and describe the methods in which we reduce the data (both real and simulated) for the purposes of our analysis. In Section \ref{sec:results}, we describe the results of our analysis. Finally, we discuss future prospects for the extension of this analysis and we conclude in Section~\ref{sec:conclusion}. 
 
\begin{center}
\begin{figure*}
\includegraphics[width=\textwidth]{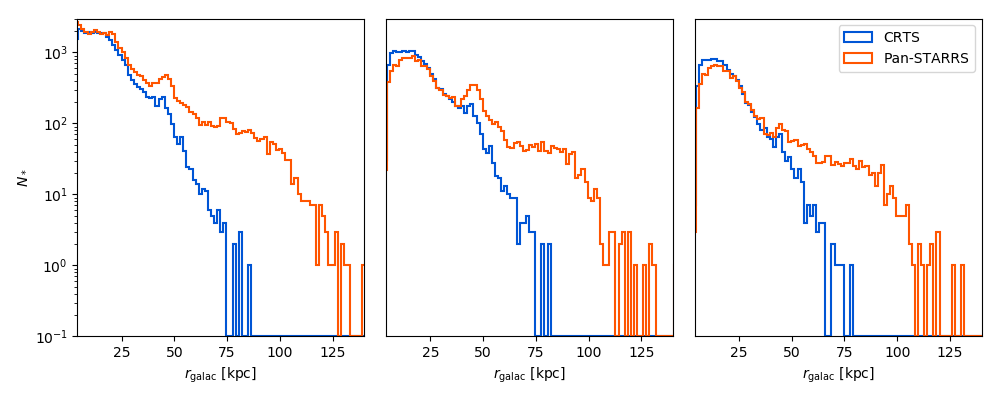}
\caption{The distribution of RR Lyrae stars with radius for our two data sets. \textit{Left Panel}: The distribution for the full data sets without any cuts. \textit{Middle and Right Panels}: The distribution after we have applied our geometry cuts (middle) and excised the Sgr stream (right). The details of the cuts are displayed in Fig.~\ref{fig:crts_fp}.}
\label{fig:galac_dist_distribution}
\end{figure*}
\end{center}

\begin{center}
\begin{figure}
\includegraphics[width=.5\textwidth]{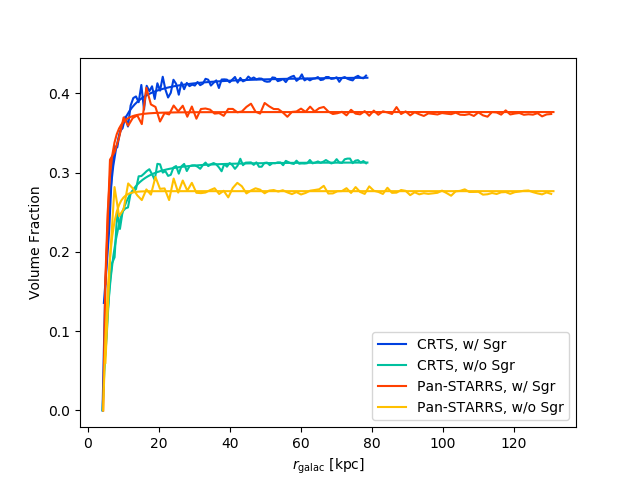}
\caption{The volume completeness of the surveys in the case of our geometry cuts with and without the Sgr stream. We calculate these volume fractions using Monte Carlo integration with $5\times 10^6$ points and then fit the resulting curves with power laws which asymptote to a fixed fraction and are set to be zero at the lower limit of the distance distribution.}\label{fig:vol_frac}
\end{figure}
\end{center}
\begin{figure*}
\includegraphics[width=.45\textwidth]{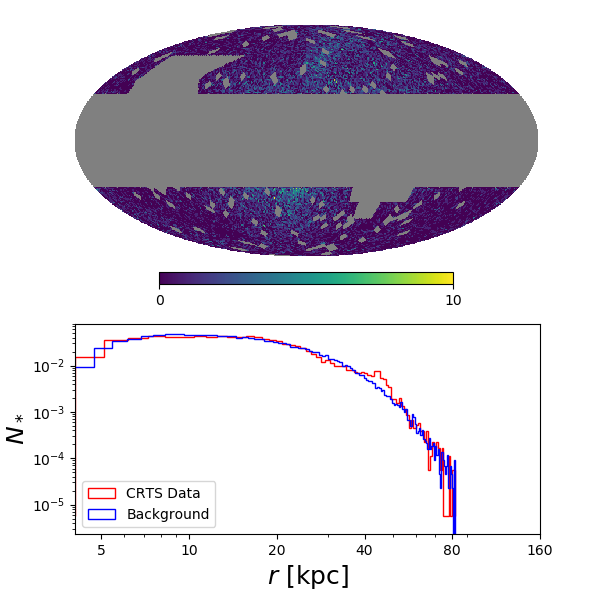}
\includegraphics[width=.45\textwidth]{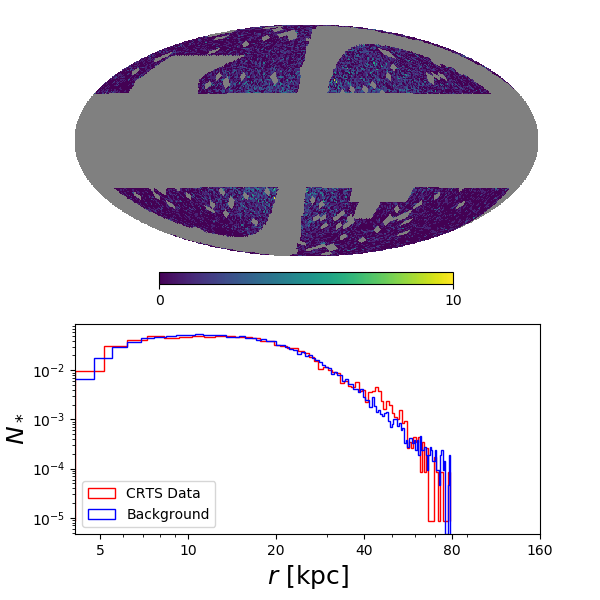}
\includegraphics[width=.45\textwidth]{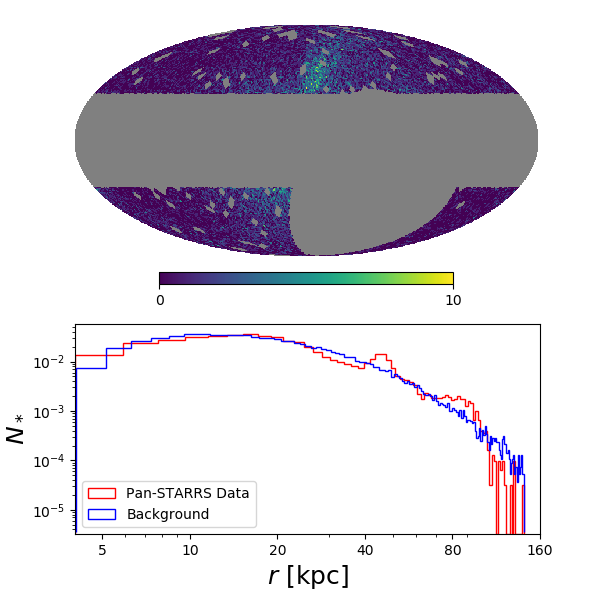}
\includegraphics[width=.45\textwidth]{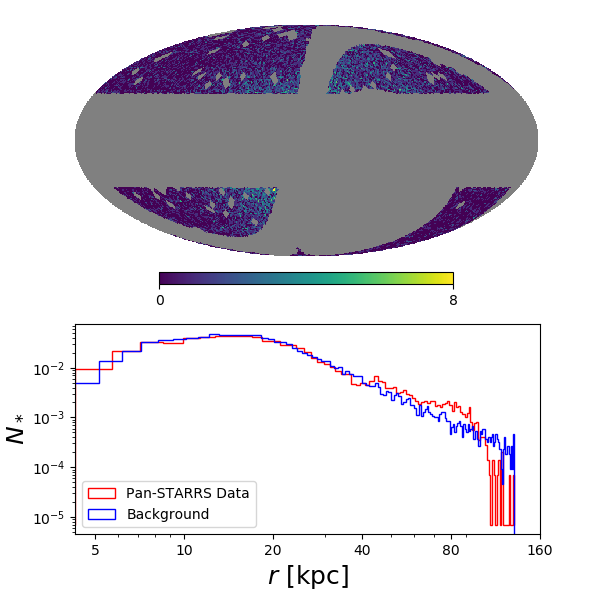}
\caption{Top Left Panels: Histograms of the distribution of the CRTS data on the sky in Galactic coordinates. Below the skyplot, we show the normalized distribution of both the CRTS data (red) and the background associated with it (blue) with distance.  We note that the background distribution matches the CRTS data quite well, except for the overdensity at $\sim 45$ kpc, which is due to the Sgr stream. Top Right Panels: The same, but for the CRTS data with the Sgr stream removed. Bottom Left and Right Panels: The same as the Top Left and Right Panels, but for the Pan-STARRS data.}\label{fig:crts_fp}
\end{figure*}

\section{Data and Methods}
\label{sec:methods}

\subsection{3D 2PCF of the Stellar Halo}
\label{sec:2pcorr}

Though many authors have used velocity space information together with 3D spatial information to quantify substructure in the stellar halo \citep[e.g.][]{Starkenburg2009,Xue2009,Xue11,Janesh2016}, this requirement is quite restrictive on the size of the dataset. At the same time, other authors have taken advantage of large datasets and simply looked at structure as it appears on the sky \citep[e.g.][]{Lemon2004, Bell2008, Deason11,Helmi2011}, but this is conversely restrictive on the amount of information extracted from the data. In choosing to look at 3D spatial structure, we work between these two approaches and ideally get the best of both worlds.

In order to quantify the substructure, we borrow a tool commonly used in Cosmology (and in several other areas such as statistical mechanics and quantum field theory): the Two-Point Correlation Function (2PCF). This statistic is of course formally defined for continuous functions or fields as the expectation value of that function multiplied by itself at a given separation over the space on which it is defined. However, when dealing with a fixed sample of data there are several commonly used estimators of this quantity.  Authors in the past \citep[e.g. ][]{Cooper2011} have used the ``standard'' estimator, defined as 
\begin{equation}
\label{eq:xi_est_standard}
\hat{\xi}_S (d) = \frac{DD(d)}{RR(d)} - 1
\end{equation}
where $DD(d)$ is the distribution of 3D distances between stars (or any point source) in the dataset and (integral to a 2PCF) $RR(d)$ is the distribution of distances between `random' points drawn from a background against which comparison is made. However, while this estimator does give a general measure of relative structure in some dataset, it does \textit{not} measure a 2PCF in a formal sense (e.g., as is seen using a simple example of 1D Gaussian data with uniform background).

For this reason, we use the Landy-Szalay estimator \citep{LS95}, which does calculate a 2PCF exactly in the limit of infinite data, and is defined as:
\begin{equation}
\label{eq:xi_est_ls}
\xi_{\rm LS}(d) = \frac{DD(d) - 2DR(d) + RR(d)}{RR(d)}
\end{equation}
where $DD(d)$ and $RR(d)$ are as defined above and $DR(d)$ is the distribution of distances between the set of data points and background points. We emphasize that hereafter we use 3D distances between the stars, not just their angular separations.

In order to formally calculate the variance of these estimators, we must evaluate the Four Point Correlation Function, which is unfortunately not computationally feasible given the size of our datasets.  Instead, we provide a lower bound on the estimates of our 2PCFs based on the propagation of Poisson uncertainty in our data.

As is evident in the definitions above, if we wish to use the 2PCF estimators on some distribution, the result depends as strongly on the background that we choose as it does on the data.

\subsection{Galactic RR Lyrae data}

Our analysis of the Galactic stellar halo substructure relies on two, complementary datasets: the Catalina Real-time Transient Survey \citep[CRTS]{crts09} and the Panoramic Survey Telescope and Rapid Response System \citep[Pan-STARRS1][]{ps1,ps2}. These separate datasets follow dissimilar footprints on the sky, have different levels of completeness in apparent magnitude, and suffer from distinct systematics. The CRTS RR Lyrae sample is a collection of five independently published datasets \citep[namely][]{Drake2013,Torrealba2015,  Drake2013_evidence,Drake2014,Drake2017}. Properties of this RR Lyrae compendium have recently been scrutinized in \citet{Belokurov2018}. The CRTS RR Lyrae sample contains 31,301 objects and probes Galactocentric radii from $\approx 2$ to 90 kpc. The completeness of the CRTS sample is a strong function of the star's apparent magnitude, but also decreases slightly around the edge of the survey's footprint. The PS1 RR Lyrae samples are presented in \citet{Sesar2017}. We require the RRab classification score to be above 0.8 to obtain 44,208 objects in total. They are distributed between Galactocentric radii $\approx 0.5$ and 150 kpc. The details of the distribution of these datasets for the various cuts are shown in Fig.~\ref{fig:galac_dist_distribution} along with the volume fraction probed by each survey as a function of radius in Fig.~\ref{fig:vol_frac}.
  
For each dataset, we make geometrical cuts on the sky in order to ensure that our measurement is not overly-influenced by completeness effects at the edges.  The cuts are shown in the panels of Fig~\ref{fig:crts_fp}.  Note that in each of these footprints, we remove areas on the sky around bound structures (Milky Way satellites, globular clusters) using a catalog compiled by \citet{torrealba18} from several different sources~\citep{mcconnachie2012,harris2010}. It may appear as though we are removing the apparent signal of structure that we wish to measure. However, our approach is to use the 2PCF for \textit{quantifying} substructure in the stellar halo, not finding it.  Compact structures, such as the ones removed, are much easier to find than the remnants of phase-mixed accretion events. In order to compare like to like, we make similar cuts in the simulations against which we compare.  We additionally use latitudinal cuts ($|b|>30^{\circ}$) to avoid problems due to incompleteness of the CRTS catalog at low Galactic latitudes. We include these cuts in the Pan-STARRS data as a means of comparison.

\subsection{Background estimate}

In order to use the 2PCF to quantify substructure in the stellar halo, we must first specify a background distribution for comparison.  We take this to be the spherical distribution of stars in the stellar halo, fit by a stepped or double power-law number density profile:
\begin{equation}
\label{eq:stepped_dens}
\nu(r) = \nu_0 \left( \frac{r}{r_0}\right)^{-\gamma} \left(1 + \left(\frac{r}{r_0} \right)^{\alpha} \right)^{\frac{\gamma - \beta}{\alpha}}
\end{equation}
where $r_0$ is the scale size, $\nu_0$ is the normalization, $\gamma$ controls the logarithmic slope of the distribution at small radii ($r \ll r_0$), $\beta$ controls that slope at large radii ($r \gg r_0$), and $\alpha$ controls the rapidity of the transition from inner to outer slope.  We then fit this profile to our data (using methods described below) and draw from the fit in order to obtain a background sample to use in our 2PCF calculation.

We begin by fitting eq.~(\ref{eq:stepped_dens}) to the radially binned data, shown as the red curves in the panels of Fig.~\ref{fig:crts_fp}. Of course, we must account for the spatial incompleteness of the data. To do this, we evaluate the goodness-of-fit of any given set of parameters for eq.~(\ref{eq:stepped_dens}) against the data by the following procedures:
\begin{itemize}
\item[1.] We evaluate the density given by these parameters on a three-dimensional grid of volume cells, centered on the Galactic center, and having a maximum extent set by that of the data. The volume cells are distributed logarithmically from the Galactic center, so as to provide better resolution at the high density inner parts of the halo. This grid consists of 100 points to a side, corresponding to $10^6$ distinct volume cells.
\item[2.] We treat the center of each volume cell as a three dimensional coordinate in space and apply the same geometry cuts on the sky (as viewed from an Earth-like position).  We are then left only with volume cells following the same geometry as our data on the sky.
\item[3.] We turn each volume cell into a mass (derived from the density at the center of the cell and the volume) and bin these masses based on Galactocentric distance using 25 logarithmically distributed radial bins. We evaluate the density of these bins as the total mass of the bins divided by the total volume of the bin.  The radius associated with the bin is then the average of the outermost and the innermost edge.
\item[4.]  We compare these binned volume cell density estimates against the data by using a quadratic interpolation as a function of radius.
\end{itemize}

In this way, we use a discretization of a continuous density profile to account for our geometrical cuts and then turn our discretization back into a continuous distribution (with which we can fit the data) using a quadratic interpolation scheme.  Using the SciPy \texttt{curve\_fit} \citep{scipy} function, we iterate this process in order to find the best fit parameters of eq.~(\ref{eq:stepped_dens}) to the data. We perform the comparison with the data by radially binning the $N_{\rm data}$ data points into $2\sqrt{N_{\rm data}}$ linearly spaced bins, again using the average of the inner and outer radial edge as the center of each bin, the number of stars in that bin divided by the volume of the bin as a density, and the square root of this number of stars (divided by the volume) as the error on that density. Since we are only accounting for the geometry of the survey on the sky, and not for the completeness at large distances, the fit distribution should not be interpreted as the true radial number density of the stellar halo. In the panels of Fig~\ref{fig:crts_fp}, we show the histogram of star counts as a function of radius, rather than the stellar density, to demonstrate that the choice of our data binning does not have any significant effect on the model parameters that are fit.

Additionally, so that we can evaluate how much of the signal is created by early-time, large scale accreted structures, we remove the largest scale structure in our dataset, namely the Sagittarius (Sgr) Stream.  We do this by excising all stars which lie within 10$^{\circ}$ of the equatorial plane of the Sgr Stream as defined in the appendix of \cite{belokurov14}. This cut is illustrated in the rightmost panels of Fig~\ref{fig:crts_fp}.

Once we have a background density and the data, we use Equation (\ref{eq:xi_est_ls}) to estimate the correlation function on any scale and at any location (subject to the limits of the data). We compute the correlation function using the software \texttt{TreeCorr}, which efficiently calculates the 2PCF (or other correlation functions) using a ball tree data structure \citep{treecorr}. \texttt{TreeCorr} needs the positions of every data point in 3-dimensional space along with any weight that we wish to assign to the data points (as we will need to do with the simulations below), the equivalent information for the background distribution, the range of scales $\Delta x$ that we would like to calculate the correlation function over, the number of desired bins over this range of scales, and the tolerance for miscalculating the correlation function (by estimating the distance between two points to be discrepant).  For all of the 2PCFs calculated here, we use the conservative tolerance of 0.001.




\begin{center}
\begin{figure*}
\includegraphics[width=\textwidth]{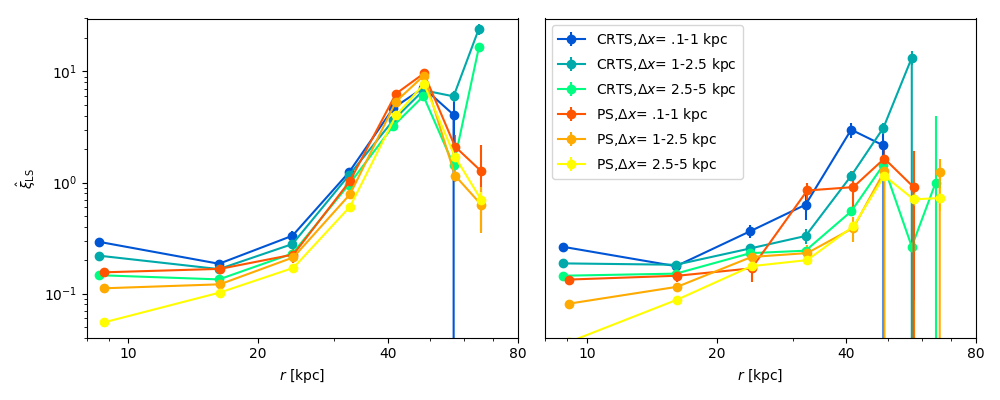}[t!]
\caption{The Landy-Szalay Estimate of the 2PCF as a function of Galactocentric radius and of the scale over which the correlation is measured. The CRTS data is indicated in cool colors, while the Pan-STARRS data is indicated in warm colors. In the left panel, we show the 2PCF estimate without removing the Sagittarius Stream, whereas in the right panel, the Sgr Stream has been removed.  The error bars on $\hat{\xi}_{LS}$ are based on simple propagation of Poisson error and represent only a lower bound on the true error in this estimate. This could account for some of the discrepancy between the two datasets. }\label{fig:survey_comp}
\end{figure*}
\end{center}

\subsection{Simulations}
\label{subsec:sims}

To compare with theoretical predictions, we use simulations of the formation of stellar halos from \citet{BandJ05}. These simulations used a merger tree formalism to construct accretion histories for each simulation in their suite and then endow each accreted progenitor with $10^5$ tracer particles \citep{BandJ05}.  This results in highly resolved accretion of low mass progenitors, but also a logarithmic mass distribution in the simulation particles that must be accounted for when computing the correlation function.

In a simulation, we are not limited by the footprint of the survey.  At the same time, we now have `simulation objects', rather than stars, as our `data points.' So, we are then naturally led to calculate the radially distributed mass density profile for each halo in the suite of simulations, and then fit eq~(\ref{eq:stepped_dens}) to this distribution. We create this distribution using all simulation objects with $r \in [1,100]$ kpc. To ensure a good fit to the zeroth order spherical structure of the halo, we remove any simulation objects which have only become unbound from their progenitors in the past 1.5 Gyrs. We then bin the $N_{\rm sim}$ simulation objects into $2\sqrt{N_{\rm sim}}$ linearly distributed distance bins.  We calculate the density in each bin by summing the mass of every particle and dividing by the total volume of the bin. We find the error in the mass $\sigma_{M_{\rm bin}}$ by writing the total mass in a bin as the product of the number of particles $N_{\rm bin}$ and the mean particle mass in the bin $\bar{m}_{\rm bin}$ and then propagating errors based on this.  We obtain:
\begin{equation}
\label{eq:bin_mass_error}
\sigma_{M_{\rm bin}} = \sqrt{N_{\rm bin}}\sqrt{ \bar{m}_{\rm bin}^2 + \sigma_m^2}
\end{equation}
where $\sigma_m$ is the standard deviation of the distribution of particle masses in that bin. 

We now have a distribution for the halo's spherical density profile that is well fit over all radii of interest.  We use this distribution to create a background for the simulations by sampling $10^6$ points from the fit distribution within the minimum and maximum radial extents of our measurement and then apply geometry cuts.  To ensure that these are roughly comparable, it is key to make sure that the footprints are somewhat similar.  So, we restrict our simulations to $r \in [4,80]$ kpc and $|b|>30^{\circ}$ in order to mimic the RR Lyrae survey data.

Since the simulations have much higher statistics than the RR Lyrae data, we compute any correlation function from the \citet{BandJ05} halos by calculating the 2PCF for 100 subsamples from the halo of interest (each with statistics similar to the RR Lyrae data) and average these answers. This makes our measurement more comparable to the RR Lyrae data (which are themselves only a sampling of the stellar halo). It is also more computationally efficient and reduces the variance in the signal due to the inclusion of massive simulation objects which dominate the mass-weighted 2PCF.

It is important to note here, that when we calculate the 2PCF for the simulations, we include a weighting for the individual simulation particles based on that particle's mass.  That way, a more massive particle contributes more to the estimate of the 2PCF than a less massive particle, accounting somewhat for the way in which each progenitor (regardless of mass) is given the same number of tracer particles in these simulations.

\begin{center}
\begin{figure}
\includegraphics[width=.5\textwidth]{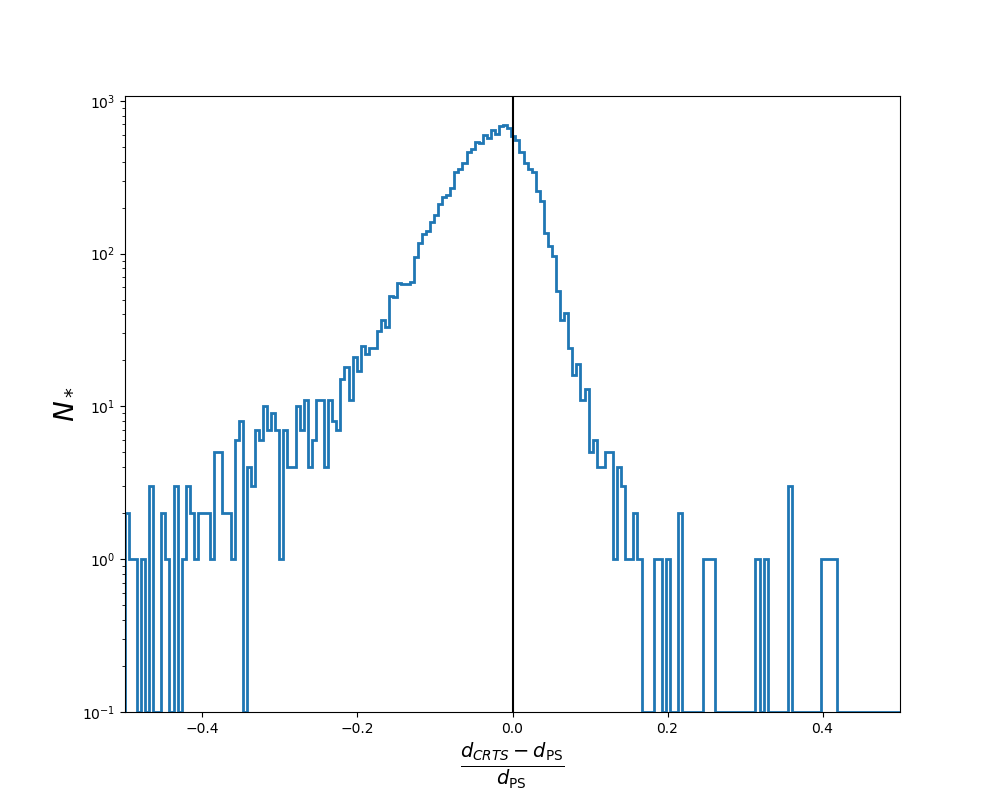}
\caption{Histogram of the percentage difference between heliocentric distances of stars from the CRTS and PS1. }\label{fig:dist_err}
\end{figure}
\end{center}

\section{Results}
\label{sec:results}

\subsection{Scale and Distance Dependence}

Our first check is a comparison of the two datasets: CRTS and PS1.  We calculate the 2PCF as a function of both Galactocentric radius and of the scale that the 2PCF probes.  We expect that the correlation function should increase as a function of radius, since at these distances there should be remnants of only partially phase-mixed substructure, given the longer dynamical mixing times. There may additionally be traces of unaccreted substructure at these distances, though we've done our best to remove these objects. We also expect that it should increase at smaller scales, since the correlation function of any smooth function increases at scales smaller than the basic scale of variation of that function.  In our data, we expect the Sgr Stream to be a significant contribution to the 2PCF at large distances. 

We show the results of this calculation in
Fig.~\ref{fig:survey_comp}. Both the trends with scale and distance
are as expected and both of the surveys are in relatively good
agreement with each other. Interestingly, the 2PCF increases
exponentially with distance out to around $r \approx 50$ kpc and then
dies off. The PS1 data, which is the more trustworthy dataset at these
distances, very clearly illustrate this drop, which we will shortly
show is also evident in many of the \citet{BandJ05} simulations.  The
fact that signal excess and the subsequent drop occurs at $r \approx
50$ kpc in the left panel of Fig.~\ref{fig:survey_comp} is simply due
to the Sgr Stream (as can be seen on comparison with the right panel),
more precisely the pile-up of the Sgr stars around the leading arm's
apo-center. The more interesting physical phenomenon is that, once
the Sgr stream is removed, there is a flattening of the amount of
structure at large distances (at least in PS1). This flattening may
indicate that the substructure (assuming that it has already been
mixed) is largely due to massive progenitors which had enough
intrinsic velocity dispersion to leave debris at these large
radii. However, the persistence of moderate growth of structure with
radius after the Sgr Stream has been removed, indicates that this
stream is by no means the only source of such substructure, as had 
been suggested in \cite{Janesh2016}.

The most notable difference between the surveys is the discrepancy of power at small radii. After cross-matching the CRTS and Pan-STARRS data sets, we found that the CRTS data systematically under-estimates the distance to the star when compared to PS1, as is illustrated in Fig.~\ref{fig:dist_err}.  This is the most likely explanation for the discrepancies in the signals at small Galactocentric distances, where the underestimate of the heliocentric distance in CRTS artificially leaks signal into the small radial bins of Fig.~\ref{fig:survey_comp}.

\begin{center}
\begin{figure}
\includegraphics[width=.5\textwidth]{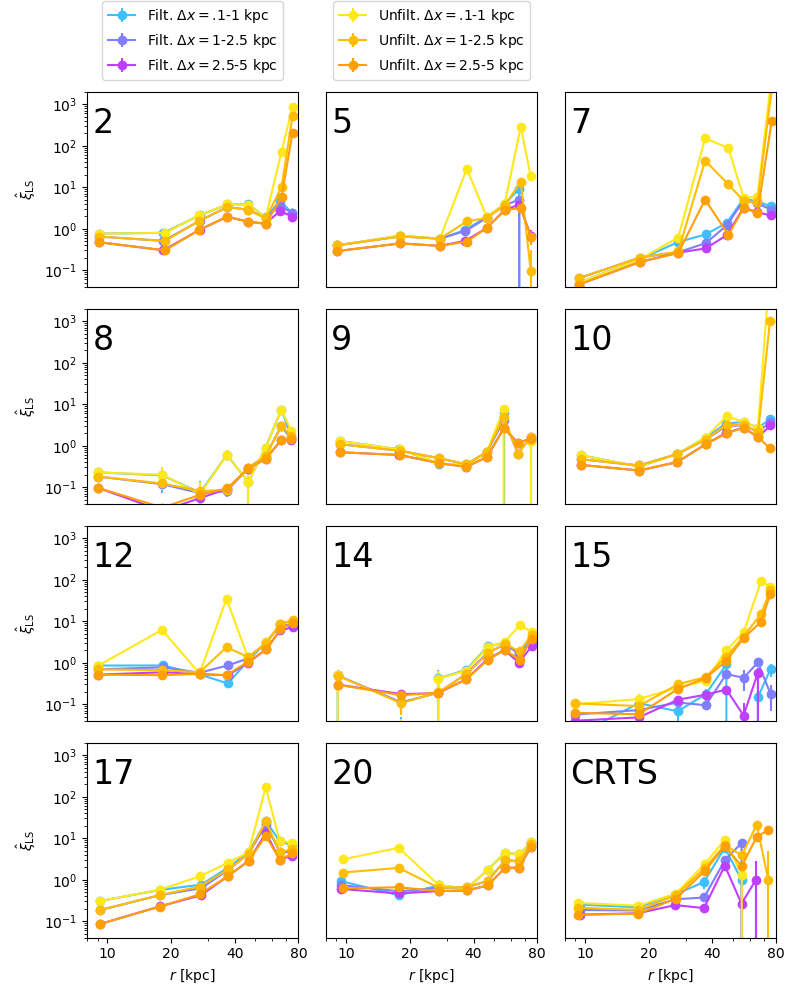}
\caption{The Landy-Szalay estimate of the 2PCF as a function of Galactocentric radius (indicated on the horizontal axis) and of the scale over which the correlation is measured. The correlation function is measured at three different scales, indicated by the different colors shown in the legend at the top of the figure.  The warm colors here correspond to the unfiltered measurement (bound structures remain for simulations, while the Sgr stream remains for the Milky Way data) while the cool colors indicate the measurement for which structure has been removed. The dataset that each panel corresponds to is indicated in the upper-left corner of each panel.  The numbers represent the Bullock and Johnston Halo ID number, whereas the last panel is the CRTS RR Lyrae data.}\label{fig:corr_all}
\end{figure}
\end{center}

Now we want to investigate the 2PCF's evolution with scale and radius in the \citet{BandJ05} simulations. Fig.~\ref{fig:corr_all} shows the Landy-Szalay estimate of the 2PCF for each \citet{BandJ05} simulation, as well as the CRTS data as a function of both distance and scale (the different colored curves). We show this calculation with and without the filtering of substructure. Structure is filtered from simulations in the manner explained in Section \ref{subsec:sims}, while the Milky Way data is filtered by removing the Sgr stream. Reassuringly, these curves are quite comparable. The expected trends in the 2PCF are shown quite clearly and on comparable scales to the Milky Way data.  This increases our confidence considerably that the \citet{BandJ05} simulations can be meaningfully compared to the data on the Milky Way.

\begin{center}
\begin{figure}
\includegraphics[width=.5\textwidth]{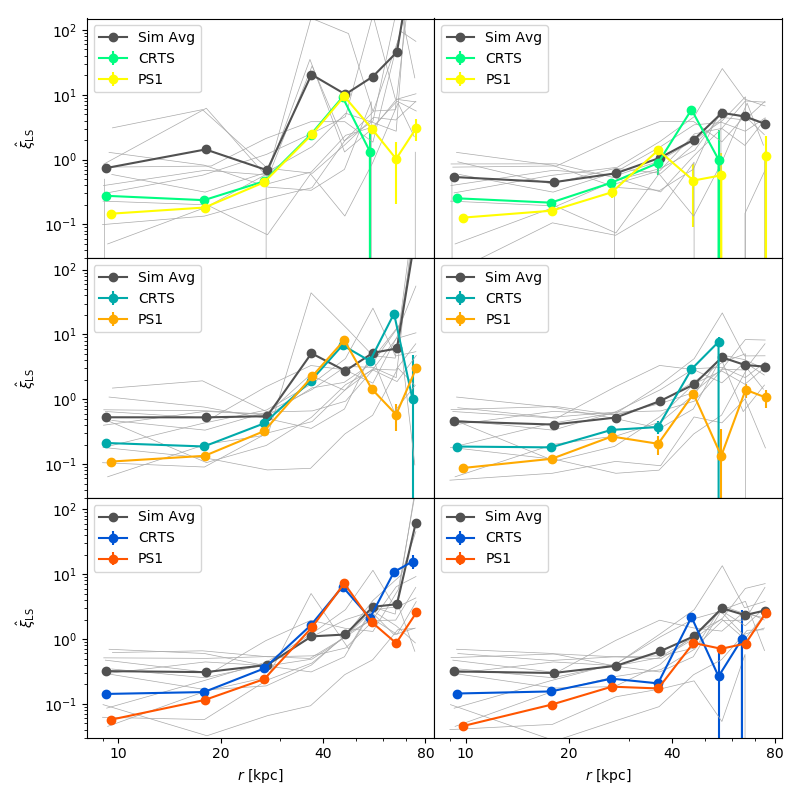}
\caption{In the left panels, we show the 2PCF as measured for halos that have had no substructure removed (Sagittarius remains for the Milky Way and recently unbound particles remain for simulations). In the right, we show 2PCFs where this filtering has been performed.  The rows of the plot move from smallest scales ($\Delta x \in [.1,1]$ kpc) at the top to largest scales ($\Delta x \in [2.5,5]$ kpc) at the bottom.  Note that the colored lines represent the survey data and correspond exactly to the colored lines shown in Fig.~\ref{fig:survey_comp}, whereas the light grey lines correspond to individual simulation curves shown in Fig.~\ref{fig:corr_all} at each scale and the dark grey line corresponds to their average. }
\label{fig:corr_all2}
\end{figure}
\end{center}

As another means of comparison, Fig.~\ref{fig:corr_all2} shows both the Pan-STARRS Data and the CRTS data on top of the \citet{BandJ05} simulations both with and without use of any filtering. This supports the idea of the rough equivalence of the filtering of the Sgr stream from the data to the filtering of recently unbound particles from the simulations. Given the compactness of some of the recently accreted substructure in the Bullock \& Johnston simulations, we can also think of the removal of this substructure as akin to the removal of the catalog of satellites that we performed on the Milky Way data. We would then expect the unfiltered halos to be widely varied in their 2PCF and have only moderate trends with scale and distance, as is indeed the case for the gray curves in the left panels of Fig.~\ref{fig:corr_all2}.

From this visualization, we can see that, while the data exhibit broad consistency with the simulations, they also exhibit a distinct deficit in structure when compared to the average simulation, especially on small scales and small Galactocentric distances.

\subsection{Accretion History}

Next we investigate differences between the CRTS/Pan-STARRS data and each of the simulations. The aim is to understand what makes some \citet{BandJ05} halos similar to the data, and hence infer properties of the accretion history of the Milky Way. Ideally, this would involve re-performing the calculations for different subsets of the accretion history of each simulated halo and observing how the removal of any part of the history changed the observed 2PCF.  However, the 2PCF is laborious to calculate and there isn't an obvious method of choosing which elements of the accretion history to leave out.  But, we can at least compare the measurements of the 2PCF for each simulated halo in Fig.~\ref{fig:corr_all2} to the 2PCF of the data and analyze the accretion histories of the halos which are most similar to the RR Lyrae data.

To make a quantitative comparison, we average the CRTS and Pan-STARRS measurements (arithmetically averaging position and 2PCF while combining errors in quadrature). We then compare against each simulation by interpolating the value of that simulations 2PCF at the center of a given CRTS/Pan-STARRS averaged radial distance bin, and computing a $\chi^2$ based on both the errors in the data, as well as in the simulations. These errors are added in quadrature as well and the simulation error is linearly interpolated between measured values. As the simulations are unlikely to be accurate for $r \lesssim 10$ kpc, we exclude this region from the calculation of our $\chi^2$ statistic, which amounts to not fitting the first of the radial distance bins.  As we wish to learn about the early-time accretion history of the Milky Way, we only include the filtered 2PCFs (right side of Fig.~\ref{fig:corr_all2}) in our $\chi^2$ statistic and create an overall $\chi^2_{\rm total}$ statistic from summing the $\chi^2$'s at each scale (the three rows of Fig.~\ref{fig:corr_all2}), i.e.,
\begin{equation}
\chi^2_{\rm total} = \chi^2_{1} + \chi^2_{2.5} + \chi^2_{5} 
\label{eq:chi}
\end{equation}
Since we are affected by incompleteness in the outer halo, we restrict our $\chi^2$ calculation to 2PCF distance bins with $r \lesssim 60$ kpc.

\begin{center}
\begin{figure}
\includegraphics[width=.5\textwidth]{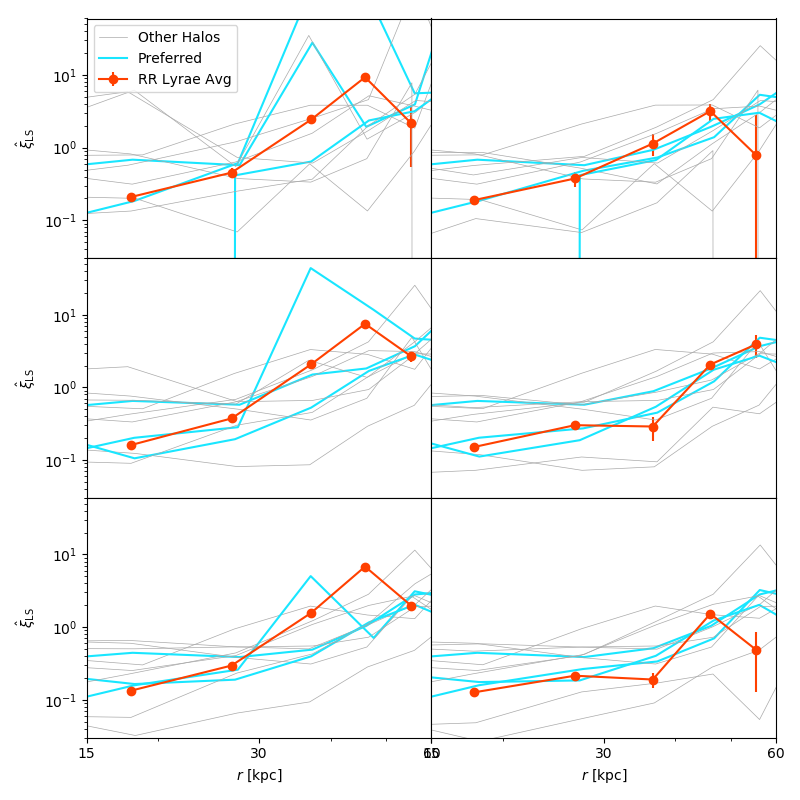}
\caption{A version of Fig.~\ref{fig:corr_all2}, where we provide the averaged CRTS and Pan-STARRS curves and high-light the simulations whose filtered curves agree best with the averaged data.}\label{fig:corr_comp}
\end{figure}
\end{center}

In Fig.~\ref{fig:corr_comp}, we show a reworking of Fig.~\ref{fig:corr_all2}, in which we have combined the CRTS and Pan-STARRS data into a single line (as it is used in the fitting) and include only those points used in the fitting. We leave out the line showing the average of the simulated halos and we highlight the 2PCFs of the three simulated halos which agree best with the data, as judged by the $\chi^2_{\rm total}$ statistic. We note again that this statistic is only calculated using the data shown on the right hand panels of Fig.~\ref{fig:corr_comp}, i.e. with most recently accreted sub-structures filtered.

\begin{center}
\begin{figure}
\includegraphics[width=.5\textwidth]{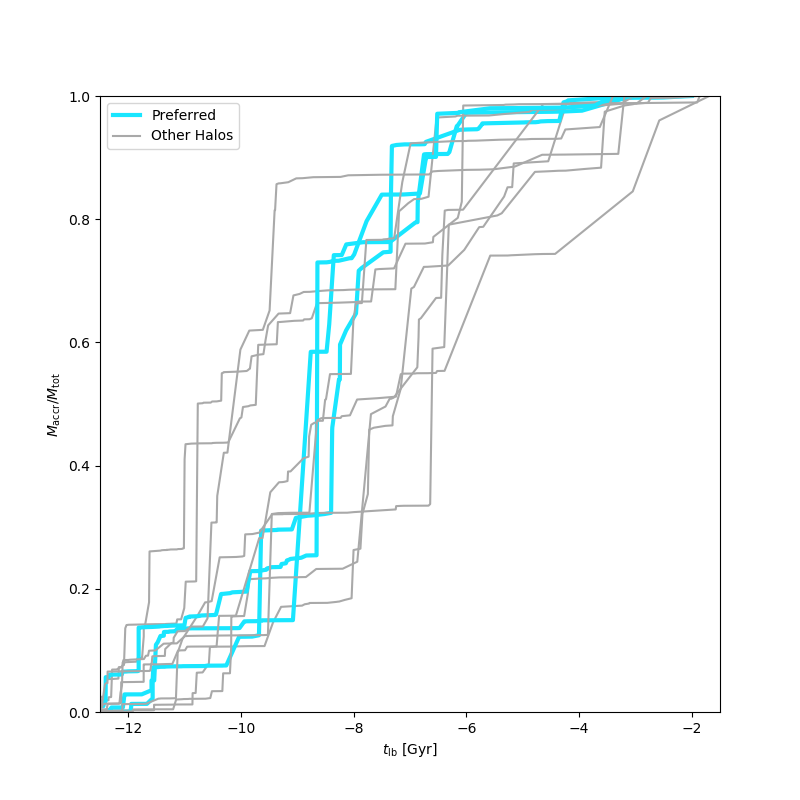}
\caption{A summary of the normalized accretion histories for each of the \citet{BandJ05} halos as a function of cosmological lookback time, $t_{\rm lb}$. At each time $t_{\rm lb}$, we plot the total stellar mass accreted up to that point, divided by the total stellar mass that is accreted up to 1.5 Gyrs ago.  In this way, we omit the `late time' accretion history. We have highlighted the accretion histories of the halos that are best fit to the RR Lyrae data, according to the metric described in eq.~(\ref{eq:chi}).  These halos are numbers 5, 7, and 14 from the \citet{BandJ05} simulations.}\label{fig:accr_hist}
\end{figure}
\end{center}

Finally, in Fig.~\ref{fig:accr_hist}, we show a summary of the
early-time accretion history of each halo in the suite of simulations
considered here. Specifically, we show the cumulative accreted mass as
a function of lookback time, which is normalized to be unity at the
right hand side of the plot, corresponding to 1.5 Gigayears ago.  We
also highlight the accretion histories of the three \citet{BandJ05}
halos which are best fit to the RR Lyrae data.  As we can clearly see,
these accretion histories are strongly consistent with one another,
and together suggest a clear preference for an accretion scenario in
which most of the early-time accretion on to the galaxy took place
around $\sim$8-9 Gigayears ago.

In fact, multiple pieces of evidence already exist that point in
concert to a massive ancient merger which delivered the bulk of the
stellar debris in the inner halo. For example, based on the
spectroscopy of local halo stars, the trend of their light element
abundances with metallicity agrees well with that observed in the
most massive Galactic neighbors such as the LMC, the SMC and the Sgr
\citep[][]{Venn2004,Tolstoy2009,deBoer2014}. Additionally, the radial
density profile of the stellar halo shows a dramatic break at around
30 kpc \citep[][]{Watkins2009,Deason11,Sesar2011}, which according to
\citet{Deason2013} could be interpreted as the last apo-centre of the
massive progenitor galaxy accreted between 8 and 10 Gyr ago. The
make-up of the halo has also been explored with simple approximations
of the stellar population tagging in \citet{Deason2015} and
\citet{Belokurov2018}. These studies concluded that the ratio of the
number of Blue Horizontal Stars to that of Blue Stragglers and the
mixture of the RR Lyrae Oosterhoff classes both indicate a small
number of old massive accretions events. Finally, \citet{GaiaSausage}
demonstrate that the shape of the velocity ellipsoid of the inner
stellar halo is inconsistent with a continuous accretion of multiple
low-mass dwarfs. According to their interpretation of the 7-D (spanned
by the phase-space coordinates and metallicity) distribution of a
large sample of main sequence stars, some two thirds of the local
stellar halo could have been deposited via the disruption of a massive
galaxy on a strongly radial orbit between redshift $z=3$ and $z=1$.

\section{Conclusions}
\label{sec:conclusion}

We have explored the utility of the Two-Point Correlation Function (2PCF) in quantifying substructure in the stellar halo. We have found a number of expected, but interesting, trends of the 2PCF -- it increases with both decreasing scale and increasing Galactocentric radius.  Our measurements of the 2PCF of RR Lyrae in the Milky Way's stellar halo are broadly consistent with the measurements of the 2PCF in the simulations of \citet{BandJ05}. However, the substructure of the stellar halo does exhibit interesting deviations from the average \citet{BandJ05} simulation at small physical scales and Galactocentric radii, where the data appears to be considerably smoother.

We used this measure of the substructure as a means of constraining the accretion history of the Milky Way.  By quantitatively comparing the measurements of substructure in the RR Lyrae data to the \citet{BandJ05}
simulations, we find that the data suggest strongly that the early-time accretion history of the Milky Way was dominated by accretion events occurring about $\sim$8 Gigayears ago and remained largely quiescent thereafter.

Looking to the future, there are several statistics that might be able to provide sharper insights into the accretion history of the Milky Way. For example, the 2PCF could be adapted to distinguish between radial and tangential accretion events. This could be accomplished by using a metric biased in either a single direction (for streams) or two directions (for shells) and then marginalizing over the orientation of these preferred directions~\citep[cf][]{Hendel15}. Another obvious extension is to include velocities, though this comes with the added problem of needing to define a metric on phase space~\citep{Starkenburg2009,Xue2009,Xue11,Cooper2010,Helmi2011}. Perhaps, due to the adiabatic invariance of actions, the 2PCF is a more powerful statistic in action space, where structure retains coherence on much longer time scales~\citep[e.g.,][]{TheBible,My18}. It may also be possible to include chemical abundance information, as this also persists on very long time scales.

To take full advantage of the various methods requires a large number of varied, detailed, high-resolution simulations of the formation of the stellar halo that go beyond the treatment of \citep{BandJ05}. With these in hand, however, we are confident that statistical tools like the 2PCF and its extensions will be crucial to providing full insight into structure formation of the Milky Way in the age of \textit{Gaia}.

\section*{Acknowledgments}  
The authors are grateful for the helpful comments of Matthew Walker,
Kathryn Johnston, David Spergel, Denis Erkal, Sergey Koposov, Eugene
Vasiliev, Alis Deason, and the members of the Cambridge Streams
Group. We would also like to thank Robyn Sanderson and Amy Secunda for
help in providing additional simulation data used in the process of
this work. The research leading to these results has received funding
from the European Research Council under the European Union's Seventh
Framework Programme (FP/2007-2013) / ERC Grant Agreement n. 308024.


\bibliographystyle{mnras}
\bibliography{halo_substructure} 


\bsp	
\label{lastpage}
\end{document}